\documentclass[pra,twocolumn,showpacs,groupedaddress,superscriptaddress,aps,10pt]{revtex4-1}
\usepackage{bm,graphicx,amsmath}
\usepackage{placeins}
\usepackage{amssymb}
\usepackage{amsmath}
\usepackage{epsfig}
\usepackage{amssymb}
\usepackage{color}
\usepackage[colorlinks,linkcolor=blue,citecolor=blue]{hyperref}
\usepackage{subfigure}

\begin{document}

\title{On the dual-cone nature of the conical refraction phenomenon}
\date{\today}

\author{A. Turpin}\email{Corresponding author: alejandro.turpin@uab.cat}
\affiliation{Departament de F\'isica, Universitat Aut\`onoma de Barcelona, Bellaterra, E-08193, Spain}
\author{Yu. Loiko}
\affiliation{Aston Institute of Photonic Technologies, School of Engineering and Applied Science Aston University, Birmingham, B4 7ET, UK}
\author{T. K. Kalkandjiev}
\affiliation{Departament de F\'isica, Universitat Aut\`onoma de Barcelona, Bellaterra, E-08193, Spain}
\affiliation{Conerefringent Optics SL, Avda. Cubelles 28, Vilanova i la Geltr\'u, E-08800, Spain}
\author{H. Tomizawa}
\affiliation{JASRI/SPring-8, 1-1-1, Kouto, Sayo, Hyogo 679-5198, Japan}
\author{J. Mompart}
\affiliation{Departament de F\'isica, Universitat Aut\`onoma de Barcelona, Bellaterra, E-08193, Spain}

\begin{abstract}
In conical refraction (CR), a focused Gaussian input beam passing through a biaxial crystal and parallel to one of the optic axes is transformed into a pair of concentric bright rings split by a dark (Poggendorff) ring at the focal plane. Here, we show the generation of a CR transverse pattern that does not present the Poggendorff fine splitting at the focal plane, i.e. it forms a single light ring. This light ring is generated from a non-homogeneously polarized input light beam obtained by using a spatially inhomogeneous polarizer that mimics the characteristic CR polarization distribution. This polarizer allows modulating the relative intensity between the two CR light cones in accordance with the recently proposed dual--cone model of the CR phenomenon. We show that the absence of interfering rings at the focal plane is caused by the selection of one of the two CR cones.

\textbf{OCIS:} (260.1440) Birefringence, (260.1180) Crystal optics, (120.5710) Refraction
\end{abstract}

\date{\today }
%
%
%
\maketitle
\section{Introduction}
\label{intro} 
\begin{figure}[htb]
\centering
\includegraphics[width=1 \columnwidth]{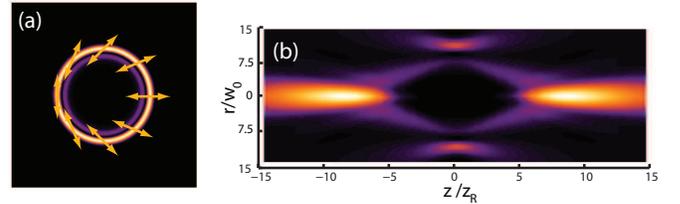}
\caption{CR intensity distribution for a circularly polarized Gaussian input beam satisfying $\rho_0 \equiv R_0 / w_0 = 10$. (a) Transverse pattern at the focal plane. Yellow double arrows indicate the polarization direction of the field along the ring. (b) Intensity distribution in the $z$--$r$ plane.}
\label{fig1}
\end{figure}

A focused Gaussian beam propagating through a biaxial crystal parallel to one of the optics axes, i.e. under conditions of conical refraction (CR), is transformed at the focal plane of the system into a pair of concentric bright rings separated by a null-intensity (Poggendorff) ring, as long as the ring radius ($R_0 = l \alpha$, where $l$ is the length of the crystal and $\alpha$ its conicity) is much larger than the waist radius of the input beam ($w_0$), i.e. for $\rho_0 \equiv \frac{R_0}{w_0} \gg 1$ \cite{Belskii,berry2004,todor,amin}, see Fig.~\ref{fig1}(a). The state of polarization at any point of the rings is linear, with the azimuth rotating so that every two diagonally opposite points are orthogonally polarized. For $\rho_0 \gg 1$, this polarization distribution only depends on the orientation of the plane of optic axes of the crystal \cite{todor,turpin_stokes}. Moving symmetrically along the $Z$ direction from the focal plane results in a more involved intensity and polarization structure and the beam forms an optical bottle. The axial spots closing the optical bottle are known as Raman spots. 

Since the last decade, CR has attracted a renewed interest as key element for a significant number of applications in fields such as optical trapping \cite{phelan2010,vault,rafailov2014,turpin_BEC_ring}, lasing \cite{laser_2014,cattoor2014}, optical communications \cite{CRFSOC_OL}, polarimetry \cite{peinado2013}, beam shaping \cite{peet_mode,turpin_hole,SGCRB}, mode conversion \cite{peet_converter} and material processing \cite{phelan:2011:oc}.

Recently, it was reported that CR can be understood in terms of the intersection of two light cones generated inside the biaxial crystal \cite{cones}.  
The vertex of each cone corresponds to one Raman spot and, in the half-way distance between them, two bright rings with Poggendorff splitting can be found due to the interference between both light cones, as shown in Fig.~\ref{fig1}. In Ref.~\cite{cones}, it was presented a proof of principle of the dual-cone nature of CR by blocking the CR rings at the focal plane with an iris (to block the outer ring) or a circular mask (to block the inner ring). The aim of the present article is to show the dual-cone nature of the CR beam 
without the need of blocking the beam anywhere and to present a self-consistent dual-cone model by taking into account both amplitude and polarization properties of the light beams.
We make use of an azimuthally segmented polarizer that mimics the CR polarization distribution to modulate the intensity of each of the two light cones, which allows us to observe their behavior separately. 
The work is organized as follows: Section \ref{theory} is devoted to explain the dual-cone theory CR. Section \ref{exp} presents our experimental results for the dual-cone nature of CR and in Section \ref{concl} we sum up our work. 

\section{Dual-cone theory of CR}
\label{theory}
Following the dual-cone model of CR, which can be obtained from the Belsky--Khapalyuk--Berry paraxial theory of CR \cite{cones}, for a uniformly polarized input light beam with axially symmetric intensity distribution, the electric field behind the biaxial crystal can be represented as a sum of two CR cones $\vec{C}_{\pm} (\rho,\varphi,Z)$:
\begin{widetext}
\begin{eqnarray}
\vec{E}(\rho,\varphi,Z) &=& \vec{C}_{+} (\rho,\varphi,Z) + \vec{C}_{-} (\rho,\varphi,Z), \label{field} \\ 
\vec{C}_{q} (\rho,\varphi,Z) &=& \sum_{q,s= \pm} A_{q s}(\rho,Z) \vec{e_s}(\varphi) (\vec{e_s}(\varphi) \cdot \vec{e_{\rm{in}}}),\label{eqC} \\
A_{qs}(\rho,Z) &=& \frac{1}{2} \int^{\infty}_{0} dk a(k) k e^{-iZ \frac{k^2}{4}} e^{iq \rho_0 k} \times \left( J_0(k \rho) -iqs J_1(k \rho) \right) , \\ \label{eqAqs}
\vec{e}_+(\varphi) &=& \left[ \cos \left( \frac{\varphi}{2} \right), \sin \left( \frac{\varphi}{2} \right) \right], \\ \label{dp}
\vec{e}_-(\varphi) &=& \left[ \sin \left( \frac{\varphi}{2} \right), -\cos \left( \frac{\varphi}{2} \right) \right].\label{dm} 
\end{eqnarray}
\end{widetext}
where dot in Eq.~(\ref{eqC}) means scalar product of two vectors. In these equations, $J_n(x)$ is n-th order Bessel function of the first kind. $a \left( k \right) = \int_{0}^{\infty} \rho E_{\rm{in}} \left( \rho \right) J_{0} \left( k \rho \right) d \rho$ is the 2D Fourier transform of a circularly symmetric input beam with polarization given by $\vec{e}_{\rm{in}} = [d_x,d_y]$. $(\rho,Z,\varphi)$ are the cylindrical coordinates normalized to the Rayleigh length and beam waist of the input beam, i.e. $Z = z/z_{\rm{R}}$ and $\rho = r/w_{0}$. $\vec{e}_{\pm}$ describe the CR polarization basis. 
For a circularly polarized input beam with helicity $\sigma=\pm$, the two CR cones $\vec{C}_{q}^{\sigma}$ ($q=\pm$) and the total CR electric field $\vec{E} = [E_x,E_y]$ can be written as follows:
\begin{widetext}
\begin{eqnarray}
\vec{C}_{q}^{\sigma} &=& \frac{e^{ i \sigma \varphi /2}}{\sqrt{2}} 
\left(A_{q+} \vec{e}_{+} - i \sigma A_{q-} \vec{e}_{-}\right), \\ \label{Eq1DualConeModelCP}
E_{x}^{\sigma} &=& \frac{e^{i \sigma \varphi / 2}}{\sqrt{2}} \left\{ \left( A_{++} + A_{-+} \right)\cos\left( \frac{\varphi}{2} \right) - i \sigma \left( A_{+-} + A_{--} \right)\sin\left( \frac{\varphi}{2} \right)  \right\}, \\ \label{Ex}
E_{y}^{\sigma} &=& \frac{e^{i \sigma \varphi / 2}}{\sqrt{2}} \left\{ \left( A_{++} + A_{-+} \right)\sin\left( \frac{\varphi}{2} \right) + i \sigma \left( A_{+-} + A_{--} \right)\cos\left( \frac{\varphi}{2} \right)  \right\}, \label{Ey}
\end{eqnarray}
\end{widetext}
where $A_{qs}$ with $q,s = \pm$ is given by Eq.~(\ref{eqAqs}). 

From Eqs.~(\ref{Eq1DualConeModelCP}) and (\ref{Ey}) it follows that each of the two CR cones can be represented as a decomposition into the two unit vectors $\vec{e}_{\pm}$ of the CR basis. Note that $\vec{e}_{\pm}$ are orthogonal to each other at any point in 3D space and they describe a non-homogeneous state of polarization. 

\begin{figure}[htb]
\centering
\includegraphics[width=0.95 \columnwidth]{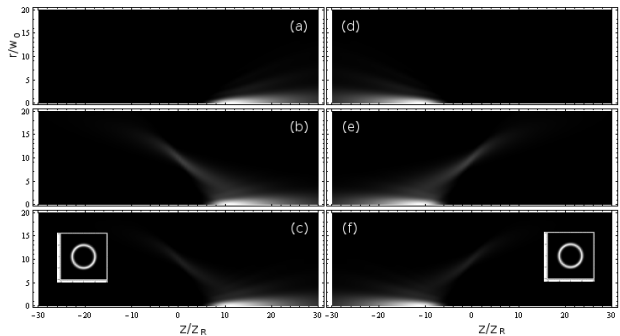}
\caption{
Decomposition amplitudes $A_{qs}$ of the CR cones $\vec{C}_{q=+}$ (a,b) and $\vec{C}_{q=-}$ (d,e) onto the CR polarization basis of $\vec{e}_{s=+}$ (a,d) and  $\vec{e}_{s=-}$ (b,e). Intensity evolution of the CR cones $\vec{C}_{+}$ and $\vec{C}_{-}$ are presented in figures (c) and (f), respectively. Their corresponding transverse profiles at the Lloyd plane are presented in the insets. $\rho_{0}=10$.
}
\label{figCRconesDecomposCRpolarizBasis}
\end{figure}

Note that in the general case, decompositions of beams on the CR polarization basis and on the CR cones are different. Each CR cone has two components in the CR polarization basis as demonstrated in Fig.~\ref{figCRconesDecomposCRpolarizBasis}. The amplitudes of this decomposition are strongly separated in space. The separation point is associated with the CR cone vertexes, i.e. the Raman spots. The $\vec{C}_{+}$ and $\vec{C}_{-}$ cones have smallest diameter at the Raman spot behind and before the Lloyd plane, respectively. Additionally, CR cones $\vec{C}_{\pm}$ have almost identical polarization distribution between the Raman spots and, consequently, they can interfere with each other, which leads to the double bright ring pattern with the Poggendorff fine splitting at the Lloyd ring plane previously observed in conical refraction. If the polarization distribution of an input beam coincides with one of the CR cones, the interference pattern at the Lloyd plane, i.e. the Poggendorff dark ring, disappears as shown theoretically in Fig.~\ref{figCRconesDecomposCRpolarizBasis}(c) and Fig.~\ref{figCRconesDecomposCRpolarizBasis}(f). Having this idea in mind and to demonstrate experimentally the dual-cone nature of the CR beam, we have designed a segmented polarizer formed by 8 sectors emulating the CR polarization -for this reason we call it as CR-polarizer- such that for two appropriate orientations of the CR-polarizer only one light ring is expected at the CR pattern at the focal plane, as shown in what follows. 

\section{Experiments}
\label{exp}

\begin{figure}[htb]
\centering
\includegraphics[width=1 \columnwidth]{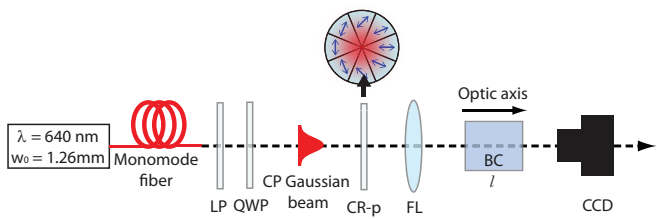}
\caption{Experimental set-up. A CR-like non-homogeneously polarized Gaussian beam is obtained by means of a segmented polarizer (CR-p). The beam is focused (FL) through the biaxial crystal (BC) and the pattern at the focal plane is recorded by a CCD camera. LP: linear polarizer. QWP: $\lambda /4$ plate.}
\label{fig_setup}
\end{figure}

Fig.~\ref{fig_setup} shows our experimental set-up. As input beam, we take a collimated linearly polarized Gaussian beam with $w_0 \approx 1\,\rm{mm}$ waist radius obtained from a $640\,\rm{nm}$ diode laser coupled to a monomode fiber. The linear polarizer (LP) and the quarter wave-plate (QWP) are used to control the polarization of the input beam and fix it to be circular. The CR-polarizer is placed after the QWP with its center coinciding with the center of the beam, which is focused by the lens ($200\,\rm{mm}$ of focal length) upon the biaxial crystal (BC). To ensure that the vertexes of the CR-polarizer are right at the center of the beam, we use an $XY$ micro-positioner. Once focused, the beam has a waist radius of $w_0 \approx 41\,\rm{\mu m}$. As BC we use a commercially available (CROptics) $\rm{KGd(WO_4)_2}$ crystal being $l = 28\,\rm{mm}$, $\alpha = 16.9$~mrad and, therefore, $R_0 \approx 475\,\rm{\mu m}$, yielding $\rho_0 \approx 11$. The polished entrance surfaces of the BC (cross-section $6 \times 4~\rm{mm^2}$) have parallelism with less than 10~arc sec of misalignment, and they are perpendicular to one of the two optic crystal axes within $1.5\,\rm{mrad}$ misalignment angle. To have a fine alignment of the BC, we mount it over a micro-positioner that allows modifying independently the $\theta$ and $\phi$ angles (when considering spherical coordinates) and we image the transverse CR pattern at the focal plane with a CCD camera, while slightly modifying the $\theta$ and $\phi$ angles until a cylindrically symmetric pattern is obtained. This means that the input beam passes exactly along one of the optic axis of the BC. The CCD camera is mounted on a translation stage to record the pattern at different planes along the beam propagation. 

\begin{figure}[htb]
\centering
\includegraphics[width=1 \columnwidth]{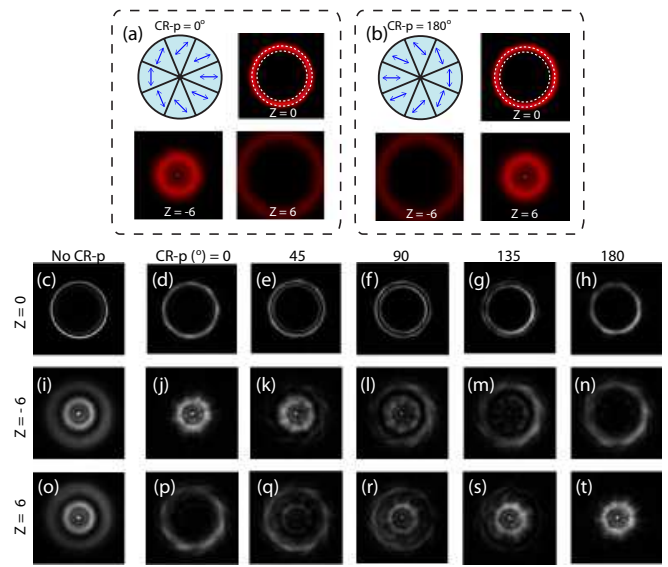}
\caption{(a) and (b) show two particular configurations of the CR-polarizer for which only one CR light ring is obtained. (c)--(h) show the transverse CR pattern at the focal plane for different orientations of the CR-polarizer, while. (i)--(n) and (o)--(t) are the corresponding transverse patterns (demagnified in size around a $30 \%$) at $Z=-6$ and $Z=6$, respectively.}
\label{fig3}
\end{figure}

The possibility to use the CR-polarizer to observe the dual-cone nature of CR is reported in Fig.~\ref{fig3}. 
Figs.~\ref{fig3} (d)--(h) show the CR pattern at the focal plane for rotation of the CR-polarizer at angles $\phi_{\rm{CR-p}} = [0\,^{\rm{\circ}},180\,^{\rm{\circ}}]$ in steps of $45\,^{\rm{\circ}}$. Fig.~\ref{fig3}(c) is the pattern obtained in the absence of the CR-polarizer. At $\phi_{\rm{CR-p}} = 0$ only one light ring is observable. As $\phi_{\rm{CR-p}}$ increases, the intensity of this ring decreases and an inner ring starts to form. At $\phi_{\rm{CR-p}} = 90\,^{\rm{\circ}}$ the pattern is clearly formed by two light rings. From $\phi_{\rm{CR-p}} = 90\,^{\rm{\circ}}$ on, the intensity of the outer ring keeps decreasing as the one from the inner increases, until having only one light ring again at $\phi_{\rm{CR-p}} = 180\,^{\rm{\circ}}$.
To provide an even clearer visualization of the effect producing by the rotation of the CR-polarizer, we have carried out the same experiment at $Z=\pm 6$, the planes where the Raman spots starts to appear. The results are shown in Figs.~\ref{fig3}(j)--(n) for $Z=-6$ and in Figs.~\ref{fig3}(p)--(t) for $Z=6$, being Figs.~\ref{fig3}(i) and (o) the pattern obtained in the absence of the CR-polarizer. At $Z=-6$ and for $\phi_{\rm{CR-p}} = 0\,^{\rm{\circ}}$, the CR pattern only has contributions from the beam center. In contrast, at $\phi_{\rm{CR-p}} = 180\,^{\rm{\circ}}$ a light ring with no intensity at its center is found. At intermediary angles of the CR-polarizer, contributions of both the central intensity and the light ring are found. At $Z=6$ the CR pattern obtained for $\phi_{\rm{CR-p}} = 0\,^{\rm{\circ}}$ is a light ring whereas at $\phi_{\rm{CR-p}} = 180\,^{\rm{\circ}}$ only intensity at the beam center is observed. 
These results, together with the one presented in Figs.~\ref{fig3}(d)--(h) indicate that the CR beam can be actually understood as two axially displaced light cones, being the focal plane a plane of mirror symmetry. 

\begin{figure}[htb]
\centering
\includegraphics[width=1 \columnwidth]{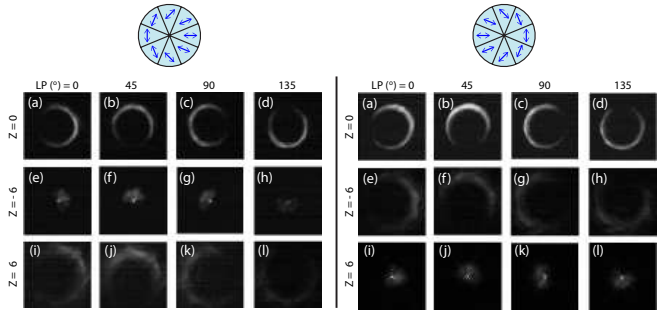}
\caption{CR transverse intensity profiles at the focal plane (first row) and at $Z=- 6$ (second row) and $Z=+6$ (third row) for orientations of a linear polarizer (LP) at angles: $[0\,^{\rm{\circ}},45\,^{\rm{\circ}},90\,^{\rm{\circ}},135\,^{\rm{\circ}}]$. Left- and right-hand side set of images correspond to $\phi_{\rm{CR-p}} = 0\,^{\rm{\circ}}$ and $\phi_{\rm{CR-p}} = 180\,^{\rm{\circ}}$, respectively.}
\label{fig_pol}
\end{figure}

It is also necessary to check the state of polarization of the light cones. Fig.~\ref{fig_pol} presents the experimental images of the CR transverse intensity profiles at the focal plane (first row) and at $Z=- 6$ (second row) and $Z=+6$ (third row) for different orientations of a linear polarizer (LP) used to analyze the state of polarization of the two cones. Left-hand set of images correspond to $\phi_{\rm{CR-p}} = 0\,^{\rm{\circ}}$ while right-hand side set of images correspond to $\phi_{\rm{CR-p}} = 180\,^{\rm{\circ}}$ see top insets. Each set of images corresponds to the $C_-$ and $C_+$ cones, respectively. The images show two remarkable features: (i) every two diametrically opposite points at the light pattern are orthogonally polarized at any plane, and (ii) the polarization distributions of both light cones are the same.

\begin{figure}[htb]
\centering
\includegraphics[width=1 \columnwidth]{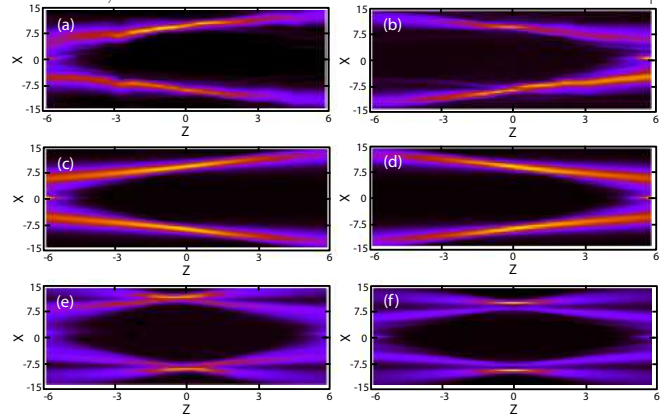}
\caption{Cuts in the $Z$--$X$ plane (where $X \equiv x/w_0$) of the beam evolution obtained for two orientations of the CR-polarizer at (a) and (c) $0^{\rm{\circ}}$, and (b) and (d) $180^{\rm{\circ}}$, showing the dual-cone nature of the CR phenomenon. (e) and (f) represent, respectively, the experimental and numerical beam propagation with the CR-polarizer being removed from the set-up. $\rho_0^{\rm{th}} = 10$, $\rho_0^{\rm{exp}} \approx 11$.}
\label{fig4}
\end{figure}

To observe the free space evolution of the light cones, Fig.~\ref{fig4} presents cuts in the $Z$--$X$ plane 
of the beam evolution of the CR pattern using the CR-polarizer. First row are the experimental results, while second row are the numerical simulations obtained by Eqs.~(6)--(\ref{Ey}). Figs.~\ref{fig4}(a) (obtained with $\phi_{\rm{CR-p}} = 0\,^{\rm{\circ}}$) and (c) correspond to the $C_-$ light cone, while Figs.~\ref{fig4}(b) (obtained with $\phi_{\rm{CR-p}} = 180\,^{\rm{\circ}}$) and (d) correspond to the $C_+$ light cone. Fig.~\ref{fig4}(e) is the experimental beam evolution with the CR-polarizer being removed and Fig.~\ref{fig4}(f) is the corresponding numerical simulation. The experimental images were taken by recording the transverse light pattern at different planes along the axial direction in steps of $5\,\rm{mm}$ and then interpolating between them using the software ImageJ. As it can be appreciated, both experiments and numerical simulations are in good agreement, confirming the dual-cone nature of the CR phenomenon. 

\section{Conclusions}
\label{concl}
In summary, we have demonstrated the dual-cone nature of the CR phenomenon by using an azimuthally segmented polarizer that mimics the CR polarization distribution to generate a non-homogeneously polarized beam. We have proved that such device allows for selecting between the two CR cones and shown the experimental free-space conical beam evolution for a Gaussian input beam. A mirror-symmetric beam evolution with respect to the focal plane for the two light cones has been obtained. In addition, we have demonstrated that the CR cones $\vec{C}_{+}$ and $\vec{C}_{-}$ have their vertexes at the Raman spot behind and before the focal plane, respectively. Finally, it has been also reported the generation of two bright rings split by the Poggendorff dark ring at the focal plane, which can be understood in terms of the interference produced by the difference on the divergence of the two co-propagating light cones \cite{cones}. 

The difficulties in the state of the art to generate high quality conical/annular light beams make the present work suitable for experiments both in atom \cite{2009_Cronin_RMP_81_1051,turpin_BEC_ring} and photophoretic trapping \cite{vault,coleman}.

\section*{Acknowledgments}
The authors gratefully acknowledge financial support through the Spanish Ministry of Science and Innovation (MICINN) (contract FIS2011-23719) and the Catalan Government (contract SGR2014-1639). A.T. acknowledges financial support from the MICINN through the grant AP2010-2310.

\end{document}